\documentclass[twocolumn]{elsart}
\usepackage{natbib}
\usepackage{graphicx}
\begin{document}
\runauthor{Taam and Ricker}
\begin{frontmatter}
\title{Common Envelope Evolution}
\author[NU]{Ronald E. Taam,}
\author[UIUC]{Paul M. Ricker}

\address[NU]{Department of Physics and Astronomy, Northwestern University, 
Evanston, IL 60208}
\address[UIUC]{Department of Astronomy, University of Illinois, Urbana, 
IL 61801}
\begin{abstract}
The common envelope phase of binary star evolution plays a central role in many 
evolutionary pathways leading to the formation of compact objects in short 
period systems. Using three dimensional hydrodynamical computations, we 
review the major features of this evolutionary phase, focusing on the 
conditions that lead to the successful ejection of the envelope and, hence, 
survival of the system as a post common envelope binary.  Future hydrodynamical 
calculations at high spatial resolution are required to delineate the regime 
in parameter space for which systems survive as compact binary systems from 
those for which the two components of the system merge into a single rapidly rotating 
star. Recent algorithmic developments will facilitate the attainment of this goal. 
\end{abstract}
\end{frontmatter}

\section{Introduction}

The evolutionary study of close binary systems with compact neutron star and 
black hole components has been a major focus of stellar X-ray astronomy ever 
since the seminal contributions by van den Heuvel and Heise \cite{vdhh}. Since
then, the discovery of compact stars in interacting binary 
systems has had an enormous impact on our basic understanding of the properties of 
these stars, on the manner in which mass is transferred from a stellar donor to 
its compact accretor, and on the evolution of these systems \cite{vvdh,tvdh}.  
The recent {\em Chandra} and {\em XMM-Newton} observations of the point X-ray 
source population in the Milky Way and in external galaxies have led to a renewed 
interest in studying the evolution of these compact systems in a variety of 
galactic environments.

Among the trademarks of the evolutionary scenarios presented in van den Heuvel 
and Heise \cite{vdhh} are the cartoons used to depict the evolution of binary 
systems to the compact stage.  These have now become the standard means of 
visualizing the possible evolutionary channels. For a recent review of the
formation and evolution of compact X-ray sources, see Tauris \& van den Heuvel 
\cite{tvdh}. Central to the construction of these channels is the existence of 
a common envelope or spiral in phase in which significant mass and orbital angular 
momentum are lost from the system. In this phase, the system is transformed from 
one of long orbital period to one of short orbital period \cite{pac}. One of 
the members of the system, a star with an evolved core in the red giant or asymptotic 
giant branch star, expands beyond the orbit of its companion.  Provided that the
stellar components are not in state of synchronous rotation at the onset of mass 
transfer, the asynchronous interaction of the two components can lead to the 
conversion of orbital energy into the kinetic energy of outflow of the common 
envelope.  As a result, it was envisioned that a main sequence-like companion 
spirals toward the core of the giant.  The common envelope is ejected leaving behind 
the progenitor remnant of the compact component with its companion at separations which are 
significantly smaller than the original radius of the giant progenitor (see reviews 
by Iben \& Livio \cite{il} for evolutions involving intermediate mass stars and 
Taam \& Sandquist \cite{ts} for evolutions involving massive stars).  Among the 
classes of binary systems for which the common envelope phase plays an 
important role are systems with neutron star (binary radio pulsars, X-ray binaries), 
black hole (X-ray sources), or white dwarf (cataclysmic variables) compact components.

In the next section, the evolution leading to the common envelope phase is described 
without reference to the evolution resulting from the collisional interactions in 
dense stellar systems \cite{ba,rs,sh}.  Of particular importance, here, are the 
conditions leading to the establishment of a non-corotating common envelope in 
primordial binary systems.  The various stages of the ensuing spiral in phase 
are presented in section 3, and the conditions leading to the successful ejection of 
the common envelope are described in section 4. Finally, we provide an overview of 
the results from hydrodynamical simulations and discuss the prospects of high 
spatial resolution calculations in the near future in the last section.

\section{Evolution to the common envelope stage}

The binary system can enter into the common envelope stage when its total systemic 
angular momentum is less than its minimum value for components in synchronous motion.
In this case, it evolves away from synchronism as a result of a tidal instability 
\cite{da,ko,co,ss,hu}. This critical limit for synchronous systems 
exists because the spin angular momentum of the more evolved and 
significantly larger stellar component can become comparable to the orbital 
angular momentum of the system.  A synchronously rotating star which evolves to 
this point cannot evolve to ever larger radii and still remain synchronized to the 
orbital motion.  Instead, the two components of the system are driven out of 
synchronism by the action of tidal forces and viscous dissipation, shrinking the 
orbit as orbital angular momentum is converted into spin angular momentum of the 
larger component of the system.  For a star of a given radius, the spin 
angular momentum ($J_{spin} \propto a^{-3/2}$) increases with decreasing orbital 
separation, $a$, while the orbital angular momentum decreases ($J_{orb} \propto 
a^{1/2}$).  Hence, for sufficiently small orbital separations, the rotational 
angular velocity of the more evolved stellar component cannot be maintained at the 
orbital angular velocity of its companion.  This evolution leading to a spiral 
in phase takes place for binary systems characterized by large mass ratios.  For 
example, systems with a red giant or asymptotic giant branch component would undergo this  
tidal instability for mass ratios greater than about 5-6.  

An alternative path to the common envelope stage can occur whenever the rate of 
mass transfer is sufficiently high such that the timescale for angular momentum 
redistribution between the spin and the orbit required to bring the components of 
the system into a state of synchronous rotation is longer than the mass transfer 
timescale. This evolutionary channel can take place when the mass transfer rate 
accelerates due to an imbalance between the variation of the stellar radius and 
the star's corresponding Roche lobe with respect to mass loss.  This mass 
transfer instability is particularly relevant to systems in which the more massive star
is characterized by a deep convective envelope.  In this case, there is a tendency 
for the donor to expand while its corresponding Roche radius decreases as mass is 
lost. The accretor is also likely to expand since it accretes at a rate faster than its 
thermal timescale and cannot assimilate the accreted mass in a state of thermal 
equilibrium. As a result,  
both components are expected to fill or exceed their respective Roche lobes.
In this pathway, systems in which the more massive star undergoes mass transfer while in the 
Hertzprung gap or in the red giant or asymptotic branch phase can enter 
the common envelope phase either via a direct or a delayed dynamical 
instability \cite{hj}. We note that evolution to the common envelope stage via a mass 
transfer instability is not limited to accreting non degenerate 
stars, but can also apply to compact accreting neutron stars \cite{kb} as well.

\begin{figure}
\begin{center}
\includegraphics*[scale=.32]{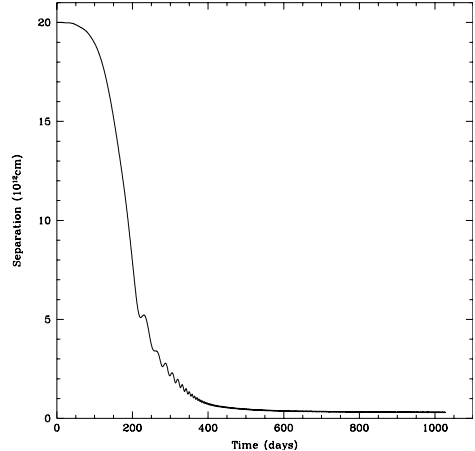}
\end{center}
\caption{The orbital separation as a function of time for a system composed of 
an asymptotic giant branch star of $3 M_{\odot}$ with a core mass of $0.7 
M_{\odot}$ synchronously rotating with a $0.4 M_{\odot}$ main sequence companion 
for a system with an initial orbital period of 0.84 yr \cite{st1}.}
\label{f1}
\end{figure}

\section{Evolutionary stages}

Substantial progress in the description of the spiral in evolution has been achieved 
from investigations of increasing complexity and realism.  Building on the early 
studies in both one \cite{tbo,mm,de} and two \cite{bt,tb1,tb2,tbr} dimensional 
modeling, the studies based on three dimensions \cite{ls,rl,ttl,tt,st1,st2} 
have provided significant 
understanding of the conditions leading to successful ejection of the common envelope.  
A particularly challenging aspect of the multi-dimensional calculations is the 
wide range of timescales and length scales that must be modelled.  In particular, 
the timescales range from hours during the late stages to years at the onset of the 
common envelope phase, and the length scales range from Earth-like 
dimensions to more than an astronomical unit. As a result, the hydrodynamical 
calculations that have been reported are primarily exploratory in nature.  
On the other hand, the existing computations have provided insight into the 
essential features introduced by the additional dimensionality considered in 
the problem and the important physical processes required for a detailed 
description underlying our understanding of this evolutionary
channel.  In the following, we provide an overview of the several distinct stages of 
the common envelope phase, focusing on the evolution during the deep spiral in phase to 
short orbital periods and the terminal phase leading to the formation of the final 
post common envelope system.  To these individual phases we now turn. 
\begin{figure}[ht]
\begin{center}
\includegraphics*[scale=.32]{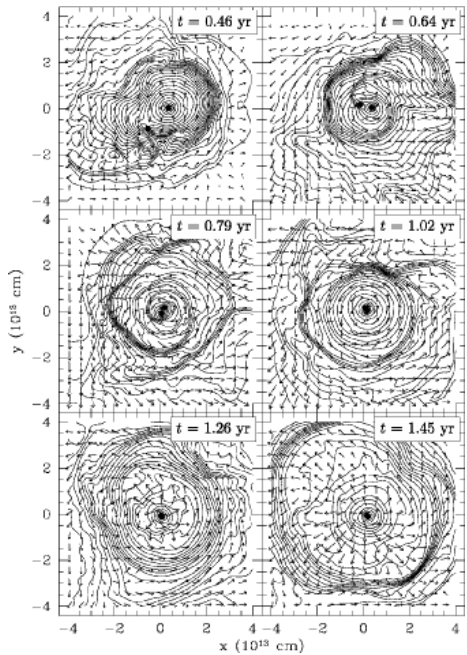}
\end{center}
\caption{The density distributions and velocity fields in the orbital plane of the 
binary described in Figure \ref{f1} \cite{st1}.  The density contours are logarithmic 
and are spaced five per decade.  The velocity vectors are scaled to the maximum in 
each panel.  The solid dots indicate the position of the cores and the distributions 
are at times given in each panel.  Spin up can be seen at a time of 1.02 yrs with 
material outflow subsequently taking place (e.g., see panels at times 1.26 and 
1.45 yrs).}
\label{f2}
\end{figure}

\subsection{Deep spiral in}

To illustrate a typical binary system, in the common envelope phase, in Figure \ref{f1} 
we show the orbital separation as a function of time for an asymptotic giant 
branch star with a low mass main sequence companion. 
It is clearly evident that the separation between the two cores decreases rapidly (on a 
timescale of about 100 days).  During the inspiral, the outer layers of the common 
envelope are tidally stripped in the orbital plane, leading to the formation of 
an outflowing circumbinary disk of material.  The relative orbit 
of the two cores becomes very eccentric during the early phase of evolution as 
most of the orbital angular momentum is converted to spin angular momentum in the
outer envelope layers.  However, as the two cores spiral closer together, 
a greater fraction of orbital energy is lost at smaller separations in comparison 
to orbital angular momentum, causing the orbit to become more circular. 

\begin{figure}[ht]
\begin{center}
\includegraphics*[scale=.32]{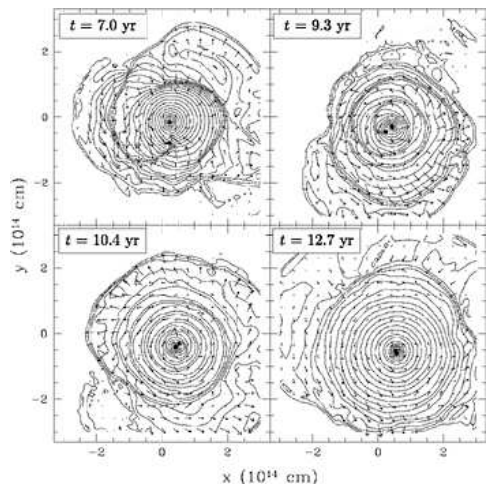}
\end{center}
\caption{The density distributions and velocity fields in the orbital plane 
of a binary composed of a $20 M_{\odot}$ red supergiant with a 1.4 $M_{\odot}$
companion at different evolution times\cite{ts}. Density contour levels 
correspond to 5 per decade, and the velocity field is scaled to 60 km s$^{-1}$ 
in all panels.  The evolutionary state of the massive star is at the
onset of core carbon burning. A tightly wound spiral emerges as the orbital
velocities of the two cores exceed the rotational velocity of the envelope gas.
The solid dots indicate the position of the cores and the initial orbital period 
is 10.6 yr. Spin up occurs at time 10.4 yr, leading to an outward flow at a time 
of 12.7 yr.}
\label{f3}
\end{figure}

The rapid orbital decay of the system causes envelope material to accelerate 
relative to the companion star, leading to the formation of shocks in the envelope. 
As the system evolves to ever smaller orbital separations, the shocks evolve into a 
tighter and tighter spiral structure pattern.  The interaction of the shocks with the 
envelope converts the orbital angular momentum of the 
two stellar components into the spin angular momentum  of the common 
envelope.  As a result, matter in 
the common envelope is spun up to nearly circular motion (see figure \ref{f2}).  
This spin up effectively causes an outward force, as centrifugal 
effects help to expand material to greater distances from the two cores. 
As shown in Figure \ref{f3}, common envelopes involving massive 
supergiant stars also show this behavior \cite{ts}.

\begin{figure}[ht]
\begin{center}
\includegraphics*[scale=.32]{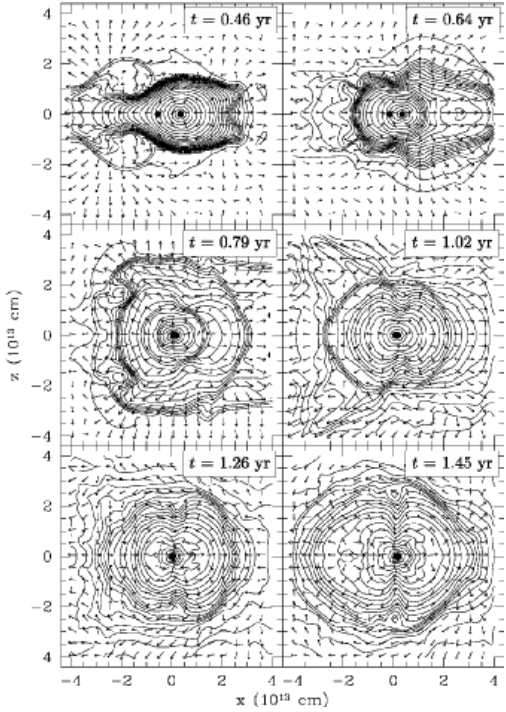}
\end{center}
\caption{The density distributions and velocity fields in the plane perpendicular to 
the orbital plane at the times for the system in Figure \ref{f2} \cite{st1}. 
Density contours and velocity vectors are as in Figure \ref{f2}.}
\label{f4}
\end{figure}

\begin{figure}[ht]
\begin{center}
\includegraphics*[scale=.35]{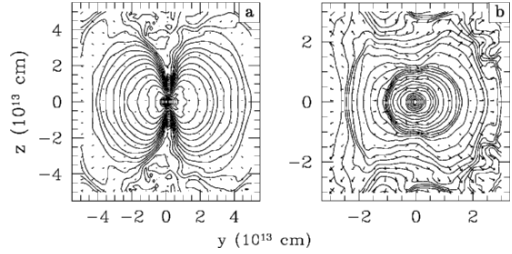}
\end{center}
\caption{The density contours and velocity field in the plane perpendicular to 
the orbital plane after the rapid infall phase \cite{st2}. Density contours are 
5 per decade, and the velocity fields in each panel are scaled to a maximum 
speed of 60 km s$^{-1}$. The mass of the red giant core and companion in the common 
envelope binary are $0.45 M_{\odot}$ and $0.35 M_{\odot}$, respectively. The 
mass of the red giant in panel (a) is $1 M_{\odot}$, and in panel (b), 
$2 M_{\odot}$.  Note that a sharper density contrast between equatorial and 
polar directions results in the case for which the mass of the companion is 
comparable to the mass in the common envelope.}
\label{f5}
\end{figure}

\subsection{Final stage}

Since the orbital energy is deposited into the common envelope on a dynamical timescale 
in the orbital plane, the ejected material is concentrated toward this plane.  The 
ejection, however, is nearly axisymmetric with respect to the angular momentum axis 
since the timescale on which the energy is deposited (comparable to the local orbital 
period) is less than the timescale of orbital decay in this stage.  The flow is not 
purely radial, as can be seen most clearly at a time of 1.45 yr in Figure 
\ref{f4}, where a circulatory flow between the equatorial plane and the polar direction 
results.  

Spin up to a greater degree is found for a red giant progenitor with a low moment of 
inertia either due to low envelope mass or relatively small radius.  In this case, 
a greater density contrast between the polar and equatorial directions can develop.  
This may be relevant to the bi-polar morphology of a class of planetary nebulae. 
Figure \ref{f5} illustrates the density and velocity distribution in 
the plane perpendicular to the orbital plane for two systems in which the mass of 
the companion and the mass of the red giant core are $0.35 M_{\odot}$ and $0.45 
M_{\odot}$ respectively.  The structure in panel (a) for a $1 M_{\odot}$ 
red giant and in panel (b) for a $2 M_{\odot}$ red giant 
illustrates the dramatic difference in the common envelope structure when the degree 
of spin up is great.  Note the evacuation of matter along the polar directions.  
Eventually an outflow is produced that is concentrated in the equatorial plane.  

\begin{figure}[ht]
\begin{center}
\includegraphics*[scale=.35]{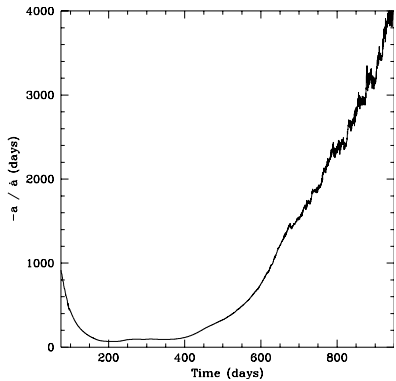}
\end{center}
\caption{The orbital decay timescale as a function of time corresponding to 
the temporal variation of the orbital separation in Figure \ref{f1} \cite{st1}.}
\label{f6}
\end{figure}

The evolution enters the terminal phase as the companion further spirals into the 
common envelope.  The orbital decay rate decelerates because most of the common 
envelope material has moved to larger radii, where its interaction with the two 
cores is greatly diminished.  As a result, the orbital decay timescale 
dramatically increases from about 100 days to more than 10 years (see Figure \ref{f6}).  

At this point we expect that the material interior to the orbit will contract 
on its local thermal timescale to the degenerate core of the giant.  Although this 
phase has not been followed, a post spiral in system should be formed with the 
companion star in close proximity to the remnant core of the former giant.  

\section{Conditions for successful ejection of the common envelope}

The hydrodynamical simulations demonstrate that material in the common envelope is 
ejected at velocities greater than that required for escape.  Although the 
calculations have not been evolved to the phase in which the entire envelope  
is lost before the two cores merge, the results to date do provide insight into 
the conditions necessary for successful ejection of the common envelope 
to take place. 

The primary requirement for ejection of the common envelope is based on energetics. 
Namely, the energy released from the orbit must exceed the binding energy of the 
common envelope from {\em both} stellar components of the system.  Generally, the
efficiency of mass ejection, as measured by the ratio of the binding energy to 
the energy lost from the orbit, is less than unity.  

Secondly, the rate at which energy is lost from the orbit must be sufficiently 
high that it can be directly converted into the energy of the outflow rather than 
being transported to the common envelope surface, where it can be radiated away.  For the 
dynamical evolution described above, the ejection process is rapid and, hence, 
adiabatic. However the efficiency of mass ejection is less than unity since 
the mass in the common envelope is preferentially ejected in the orbital plane of 
the binary system at velocities which are greater than the escape speed.  

Finally, the timescale on which the mass is lost from the system must be shorter
than the inspiral timescale of the binary, for otherwise the two cores will spiral 
together and merge before the entire common envelope is ejected. 

\begin{figure}[ht]
\begin{center}
\includegraphics*[scale=.3]{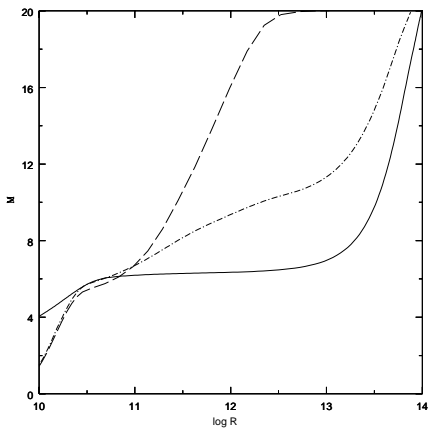}
\end{center}
\caption{The variation of mass (in $M_{\odot}$) with radius for a $20 M_{\odot}$ 
red supergiant. The dashed, dashed dot, and solid curves correspond to a 
central helium content of 0.5, 0.34, and 0 respectively \cite{ts}. Note the extensive
region between $5 \times 10^{10}$ and  $5 \times 10^{12}$ cm in which the mass varies 
very gradually with radius at the stage of carbon core burning.}
\label{f7}
\end{figure}

\begin{figure}[ht]
\begin{center}
\includegraphics*[scale=.3]{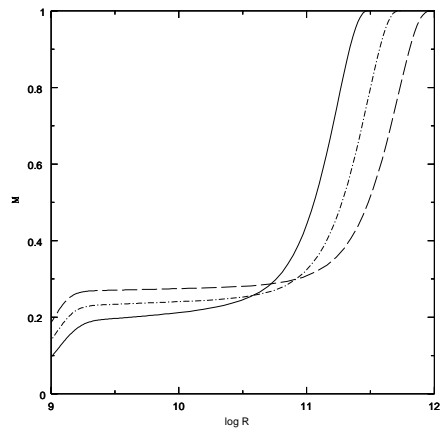}
\end{center}
\caption{The mass profile for a $1 M_{\odot}$ red giant star in different phases of 
evolution \cite{st2}. The solid, dashed-dot, and dashed curves denote evolutionary 
phases in which the mass of the degenerate helium core is 0.19, 0.23, and 0.27 $M_{\odot}$
respectively.}
\label{f8}
\end{figure}

This latter condition 
points to the importance of a core envelope structure characteristic of evolved 
stars on the giant branch. For such stars, the density profile above the nuclear 
burning shells is steep so that the mass enclosed in an extensive region is 
small.  Examples of this particular structure are illustrated in Figure \ref{f7} 
for a $20 M_{\odot}$ star in different stages of core helium 
burning and in Figure \ref{f8} for a $1 M_{\odot}$ red giant star characterized 
by a range of degenerate helium core masses. 

In both figures an extensive region containing little mass is present. 
The flat mass profile naturally develops in the 
late core helium burning stage of a massive star when the central helium content 
nearly vanishes.  For more advanced evolutionary states, the size of this region 
increases.  Similarly, red giant stars show such profiles for helium degenerate 
core masses greater than about $0.2 M_{\odot}$, extending beyond a solar radius for 
helium core masses greater than about $0.27 M_{\odot}$. 

In cases for which the loss of orbital energy does not lead directly 
to hydrodynamical ejection of the deep layers of the common envelope, mass can 
still be lost, but on a timescale significantly greater than the dynamical timescale. 
Here, the rate of energy loss from the orbit supplements the stellar luminosity 
generated by the nuclear burning shells, enhancing the rate of mass loss via the
radiative processes responsible for stellar winds. Such a circumstance may lead to 
the ejection of the common envelope, especially 
when the envelope is sufficiently spun up. Detailed calculations reveal that 
such conditions occur when the mass of the common envelope is comparable to the 
mass of the inspiraling component. This outcome for the common envelope phase may 
be related to an alternative prescription for common envelope evolution based on 
angular momentum arguments under the implicit assumption that energy is conserved. 
\cite{nt}.

Independent of whether the final phase of the common envelope is described as 
a dynamical or more gradual event, the successful ejection of the envelope requires 
progenitor stars in the red giant or asymptotic giant phase.  For progenitors on the 
main sequence or in the Hertzsprung gap, it is likely that the binary components 
will merge in the process, leaving behind a single rapidly rotating star, with possibly 
a circumstellar disk.  The class of rapidly rotating FK Comae giants \cite{bs} may 
be an observational manifestation of such a merger event.

\section{Summary and future work}

Computational hydrodynamical investigations of the common envelope phase have led 
to a number of results with significant bearing on the outcome of the post spiral 
in state of the binary system.  

\subsection{Overview of results}
 
\begin{itemize}
\item The binary orbit decays rapidly after the onset of the common envelope phase,
occurring on a dynamical timescale (i.e., on a timescale comparable to the orbital 
period for which the system entered the common envelope phase).
\item As a result of the effectiveness of the gravitational torques in removing 
orbital angular momentum from the system, the orbit of the binary system becomes 
eccentric.  Circularization of the binary orbit takes place in the deep layers 
of the common envelope as a significant fraction of the orbital energy is lost 
during this phase. 
\item There is strong evidence for significant spin up of gas surrounding the region 
containing the core of the giant progenitor and the inspiralling companion, reducing 
the effective gravity and making mass ejection easier. 
\item Matter is ejected from the common envelope in all directions, with a 
preference for the orbital plane of the binary system, producing a density 
contrast in the ejecta between the 
equatorial and polar directions. 
\item The efficiency of the mass ejection process, as measured by the ratio of the 
binding energy of the envelope with respect to the binary system to the orbital 
energy released from the binary during the common envelope phase, is less than about 
40-50\%.
\item The stabilization of the orbit at small separations during the final phase of 
the common envelope is easiest for progenitor stars which have steep 
density gradients above the nuclearly evolved core (so that the mass profile 
is flat) and/or where the mass of the common envelope is comparable to the mass of 
the inspiralling companion. 
\end{itemize}

Based on these results, one finds that the survival of the remnant binary as 
a post spiral in system resulting from the successful ejection of the common 
envelope is favored for systems in which the giant-like progenitor star is 
characterized by larger ratios of the core mass to total mass (roughly greater 
than about 0.2).  Thus, formation of cataclysmic variable type systems is expected
to be viable for red giant and asymptotic giant star progenitors with core masses 
greater than about $0.2 M_{\odot}$ and $0.6 M_{\odot}$ respectively.  For more 
massive stellar progenitors (i.e., for stars more massive than about $12 M_{\odot}$) 
relevant to the formation of low mass X-ray binary and intermediate mass X-ray 
binary systems, ejection of the common envelope is favored for advanced evolutionary 
stages of the progenitor star during its late core helium burning stage and beyond. 
We note that for stars more massive than about $40 M_{\odot}$ stellar winds may 
significantly affect their evolution, precluding the Roche lobe overflow and mass 
transfer phase since such stars do not enter the red supergiant phase \cite{vd}. 
For systems entering the common envelope phase, significant shrinkage of the 
binary orbit is possible, leading to a reduction of the orbital separation by more 
than a factor of 100 from its initial value. 

\subsection{Ongoing work}
  
Although the calculations that have been carried out in previous studies have 
provided much insight into the phases of the common envelope stage, calculations 
at high spatial resolution will be necessary to further quantify the outcome 
of the common envelope phase.  In recent years the development of sophisticated 
computer methodologies has made it possible to achieve this goal.  Specifically, 
adaptive mesh refinement techniques have advanced to the point where such 
calculations can now be envisioned.  For a recent review see Norman \cite{no}. 
These methods allow one to maintain 
high spatial resolution in the core regions as the two cores spiral in together. 

\begin{figure}[ht]
\begin{center}
\includegraphics*[scale=.3]{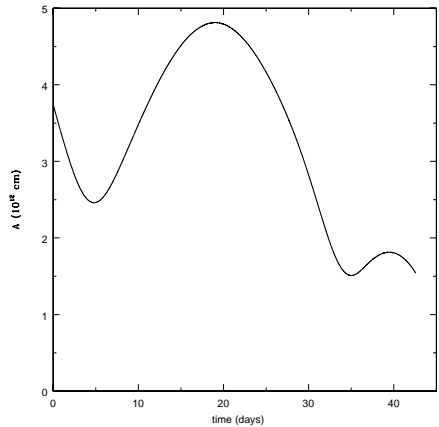}
\end{center}
\caption{The orbital separation as a function of time for a system composed of 
a red giant branch star of $1 M_{\odot}$ with a $0.7 M_{\odot}$ main sequence 
companion in which the initial orbital period is 1 month.}
\label{f9}
\end{figure}

Recently, we have initiated high resolution calculations using such techniques.
Since the core of the giant and its companion are both too compact to resolve 
even with an adaptive mesh, and because of their much higher densities with respect
to the matter in the common envelope, they couple to the gas primarily via their 
gravitational fields.  Therefore we have modelled the giant core and its 
companion as particles using an $N$ body solver based on the particle mesh method 
\cite{he}.  The companion is represented by a single particle, and the 
giant core is represented by a uniform, spherical cloud of $2\times10^5$ 
particles with core radius three times the smallest zone spacing. All of the particles
in the cloud move rigidly together with the cloud's center of mass. This 
arrangement ensures that the mapping of particle densities onto the mesh and of 
the gravitational forces onto the cloud's center of mass are free of Cartesian grid 
effects introduced by the use of a cloud-in-cell mapping kernel.
An adaptive multigrid solver is used to solve the Poisson equation for the gravitational
potential \cite{br,mc}.  Multigrid algorithms accelerate the convergence of relaxation 
methods by covering the computational domain with a hierarchy of meshes with different 
spacings and applying relaxation to each mesh.  The single-mesh convergence rate is 
controlled by error modes having the longest wavelengths compared with the mesh spacing. 
Thus the longest wavelengths in the domain converge most rapidly on a very coarse mesh. 
By combining results from all meshes a multigrid algorithm brings all wavelengths into 
convergence at the same rate. 

\begin{figure*}
\begin{center}
\includegraphics*[scale=.70]{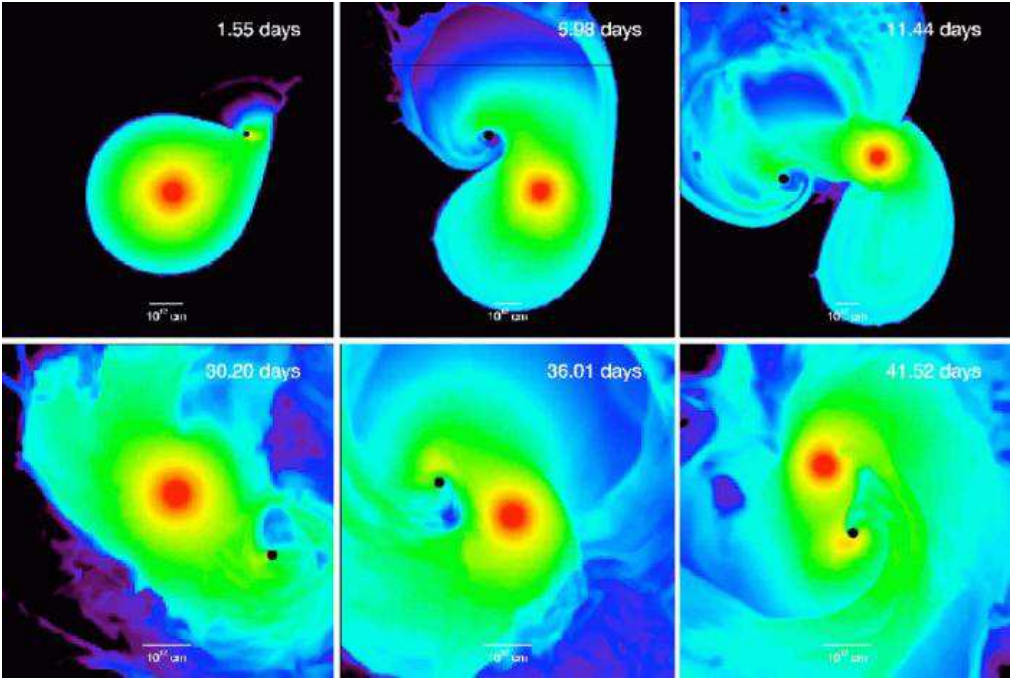}
\end{center}
\caption{The density distribution in the orbital plane during the initial and 
spiral in phase of a binary system composed of a red giant branch star of 
$1 M_{\odot}$ with a $0.7 M_{\odot}$ main sequence companion for a system with 
an initial orbital period of 1 month. Each of the six panels provides the 
evolution time, which ranges from 1.55 days to 41.52 days.  For convenience, 
a scale of $10^{12}$ cm is also indicated. The black dot indicates the position 
of the red giant companion.}
\label{f10}
\end{figure*}

The use of the adaptive mesh refinement technique and the multigrid 
method for the gravitational potentials improves on our 
previous studies \cite{st1,st2}, allowing calculations not limited to systems 
of extreme mass ratio, in which the center of mass moves only slightly during the evolution. 
We have performed several preliminary calculations of the common 
envelope phase using these techniques in the  FLASH code \cite{fr}. 
FLASH is an adaptive, parallel, 
multi-dimensional hydrodynamical simulation code in which the equations of 
compressible gas dynamics are solved using the Piecewise-Parabolic Method \cite{cw}
on a structured, rectangular grid.  PPM  
was developed to provide very accurate solutions for flows containing 
sharp discontinuities, and it greatly improves upon the hydrodynamics 
solver used in the nested grid simulations.

As an example, we illustrate the common envelope evolution of a binary system 
consisting of a $1 M_{\odot}$ red giant progenitor with a $0.7 M_{\odot}$ main 
sequence companion at an evolutionary phase when the dynamical evolution 
occurs on a timescale of about 1 month.  The orbital separation of the two 
cores within the common envelope binary system is illustrated as a function of time 
in Figure \ref{f9}.  For the case in which the red giant was rotating at half 
the synchronous rate, the orbit rapidly decayed by a factor of 2 within about 1 
month, with the relative orbit passing through two phases of apastron and periastron 
passage. 
 
In Figure \ref{f10}, we show the early phases of the spiral in process calculated 
using FLASH at three time slices at the onset of the evolution (upper 3 panels) 
and toward the end of the calculation (lower 3 panels) in the innermost $5 \times 
10^{12}$ cm of the binary. The evolution is 
similar to that described in Section 3, with the development of spiral shock waves 
emanating from the two cores.  However,  most importantly, it is 
seen that the adaptive mesh refinement technique performs very well in resolving 
the vicinity of the red giant core and the outer layers of the common envelope.  
We remark that such an evolution could not have been simulated using our previous 
stationary nested grid technique due 
to the limitations on the mass of the companion star which, for sufficiently large 
masses, caused the core of the red giant to move significantly off the 
highest spatially resolved grid.  

Evolutionary calculations such as these in which the innermost regions of the common 
envelope surrounding the two cores are highly resolved will be necessary to 
determine whether the binary orbit stabilizes or continues to shrink.  Such 
investigations are essential for determining the dependence of the efficiency 
of the mass ejection process on the mass of each component, their evolutionary 
stage, and orbital separation.  Using our model binary systems we plan to  
systematically investigate the mass ejection process as a function of these system 
parameters. This is critically important since the generality of the hydrodynamical 
results with regard to binary system parameters must be ascertained. As two possible 
mechanisms have been identified for the emergence of the system from the common 
envelope phase, an exploration of their different dependencies on the binary system 
parameters is required before definitive conclusions can be drawn for use in 
population synthesis calculations.  This study will allow us to distinguish 
the regions in progenitor parameter space for which 
the post common envelope systems survive from that in which the systems merge 
into a single rapidly rotating object.  Such results will 
provide important input for population synthesis modeling, allowing us to 
determine the birth rates for the formation 
of compact systems in a variety of galactic environments (metal rich star burst 
regions vs. metal poor regions). The comparison of such 
rates with observations 
will allow us to establish the evolutionary channels for these celestial sources. 

\ack{This work was partially supported by the National Center for Supercomputing 
Applications under grant AST040024 and utilized the NCSA Xeon Linux Cluster.  Partial 
support has also been provided by the NSF through grant AST-0200876. The software 
used in the FLASH code was developed by the DOE-supported ASC/Alliance Center 
for Astrophysical Thermonuclear Flashes at the University of Chicago.}

\end{document}